# Anisotropic Optical Properties of Layered Germanium Sulfide

*Dezhi Tan,*,† Hong En Lim,† Feijiu Wang,† Nur Baizura Mohamed,† Shinichiro Mouri,† Koirala Sandhaya,† Wenjing Zhang,† Yuhei Miyauchi,† Mari Ohfuchi,‡ and Kazunari Matsuda*,†*

†*Institute of Advanced Energy, Kyoto University, Uji, Kyoto 611-0011, Japan*

‡*Fujitsu Laboratories Ltd., Atsugi 243-0197, Japan*

Abstract

Two-dimensional (2D) layered materials, transition metal dichalcogenides and black phosphorus, have attracted much interest from the viewpoints of fundamental physics and device applications. The establishment of new functionalities in anisotropic layered 2D materials is a challenging but rewarding frontier, owing to their remarkable optical properties and prospects for new devices. Here, we report the anisotropic optical properties of layered 2D monochalcogenide of germanium sulfide (GeS). Three Raman scattering peaks corresponding to the $B_{3g}$, $A_g^1$, and $A_g^2$ modes with strong polarization dependence are demonstrated in the GeS flakes, which validates polarized Raman spectroscopy as an effective method for identifying the crystal orientation of anisotropic layered GeS. Photoluminescence (PL) is observed with a peak at around 1.66 eV that originates from the direct optical transition in GeS at room temperature. Moreover, determination of the polarization dependent characteristics of the PL and absorption reveals an anisotropic optical transition near the band edge of GeS, which is also supported by the density functional theory calculations. This anisotropic layered GeS presents the opportunities for the discovery of new physical phenomena and will find applications that exploit its anisotropic properties.

Layered materials, such as graphene and transition-metal dichalcogenides (TMDs, e.g., $MoS_2$ and $WSe_2$) that are built up of weak van der Waals bonded layers can be mechanically exfoliated to form stable two-dimensional (2D) units of atomic thickness, which show remarkable mechanical, electronic, and optical properties.[1-4] The demonstration of a rich variety of physical behaviors and applications suggests that these layered materials are a new platform for scientific and technological purposes. Inspired by these exciting achievements, a new group of emerging layered materials with low crystal symmetry (e.g., black phosphorus (BP) and mono-chalcogenides), has attracted increasing attention because of their unusual electrical and optical properties. In contrast to graphene and typical TMDs, the physical properties, such as electrical mobility, photoluminescence (PL), photoresponsivity, and thermoelectric property, in the rediscovered low-symmetry layered materials exhibit a strong dependence on the in-plane crystal orientation, which brings new elements to the study of 2D materials. For example, the puckered and anisotropic crystal structure in BP causes significant anisotropic electronic states along the zigzag and armchair directions, which lead to intriguing anisotropic carrier transport and optical properties.[1,5-7] Owing to these demonstrations and the expected superior electronic, spintronic, and photonic phenomena in the 2D monolayer case, increasing attention has been focused on exploring new members in this layered material family, especially those with an anisotropic structure.[1,5,6,8-10]

The 2D layered monochalcogenides (MX; M = Ge, Sn and X = Se, S) have similar anisotropic crystal structures to BP, which are new systems for the study of the anisotropic physical properties of 2D layered materials. Among these 2D layered monochalcogenides, germanium sulfide (GeS) is considered to be a promising material for high-efficiency solar cells and photodetectors with high sensitivity and external quantum efficiency.[11,12] The anisotropy along the armchair and zigzag directions in the crystal structure suggests that GeS will exhibit interesting anisotropic physical properties.[13-15] The thickness dependent optical properties, high carrier mobility, and large excitonic effects in GeS have also been predicted by theoretical calculations.[14,16,17] Moreover, a large spin-orbit interaction is expected, in contrast to the case of BP.[13] GeS also exhibits superior stability of structure and electronic properties in monolayer form than BP.[17,18] However, existing experimental studies of the optical properties of the 2D layered GeS are insufficient and there are still no reports of its anisotropic PL properties or detailed investigation of anisotropic Raman scattering spectroscopy.[15,19] Such studies are needed to obtain a better understanding of the optical properties of GeS. In addition, a better understanding of the anisotropic Raman scattering signals of GeS will also provide a simple but effective and precise method to identify its crystal orientation, which is an essential precondition to the study of the anisotropic properties and the fabrication of desirably oriented devices for interesting optoelectronic applications.[5,6,20,21]

In this paper, we explored the anisotropic optical properties of GeS flakes by using polarization dependent Raman scattering, absorption, and PL spectroscopies. Raman scattering peaks with strong polarization anisotropy were observed, which are useful for determining the crystal direction of GeS. We also observed a significant PL peak near the optical band edge at around 1.66 eV whose intensity is dependent on the excitation polarization. Optical absorption spectroscopy and density-functional theory (DFT) energy-band calculations suggest that the PL mainly arises from the direct optical transition in GeS. Furthermore, polarized absorption spectroscopy and the DFT calculations indicate that the anisotropic PL originates from an anisotropic optical transition near the band edge. The revealed unique anisotropic nature of GeS may lead to unprecedented possibilities for the creation of novel electronic and optoelectronic devices. Our observation and discussions are also of general significance in understanding the optical properties of 2D layered materials with in-plane anisotropy.

**RESULTS AND DISCUSSION**

Layered GeS has a distorted orthorhombic structure (space group Pcmn-$D_{2h}^{16}$), as shown in Fig. 1a. The puckered honeycomb lattice of GeS, similar to that in BP, has an anisotropic crystal structure along the armchair and zigzag directions (X and Y directions, defined in Fig. 1a). Figs. 1b and 1c show optical and atomic force microscope (AFM) images of a typical layered GeS flake that was prepared by the mechanical exfoliation method and transferred onto the SiO$_2$/Si substrate. The optical image shows a clear cleaved surface of GeS due to the layered structure formed by weak van der Waals interaction. The atomically flat surface is confirmed by the AFM image of the GeS flake on a SiO$_2$/Si substrate, and the thickness of this flake evaluated from the height profile in the inset of Fig. 1c is about 65 nm.

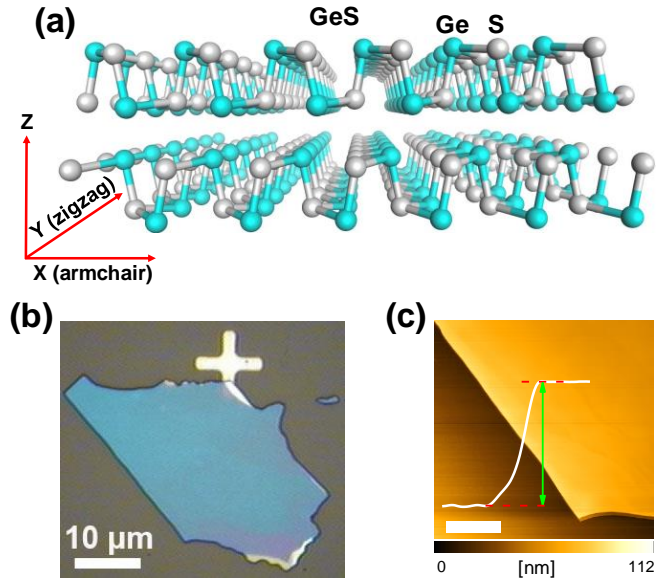

**Fig. 1.** (a) Schematic crystal structure, (b) Optical image and (c) AFM image of a 65 nm-thick GeS flake. The scale bar in AFM image is 5 μm.

Raman scattering spectroscopy is a very useful tool for the study of material structure and properties and is well-established for characterizing the lattice vibrations (phonons) and lattice dynamics of layered materials.[22-25] The Raman scattering spectrum of a 100 nm-thick GeS flake in the lowest panel of Fig. 2a displays three clear peaks at around 215.3, 240.8, and 271.8 cm$^{-1}$, which can be assigned to three phonon modes with $B_{3g}$, $A_g^1$, and $A_g^2$ symmetry, respectively.[26,27] The $B_{3g}$ and $A_g$ modes correspond to the in-plane shear vibration of adjacent layers parallel to one another in the Y (zigzag) and X (armchair) directions, respectively.[26-28] We have also prepared GeS samples with difference thickness. The Raman spectra for the GeS flakes with thickness of 8, 14, 20, and 27 nm are shown in Fig. S1. The Raman spectra do not drastically change above 8 nm, which suggest that the structure of GeS does not change.

Polarized Raman scattering spectroscopy has been widely used for determining the crystal structure anisotropy of layered materials.[24,25] We conducted polarized Raman spectroscopy for a GeS flake under parallel (//) and cross-polarization (⊥) configurations. In the case of parallel (cross-polarization) configuration, the incident light polarization (excitation side, $e_i$) was parallel (perpendicular) to the scattered light polarization (detecting side, $e_s$), which were both controlled by inserting linear polarizers into the optical path. The polarized Raman spectra were obtained by rotating the sample to change the angle between the crystal direction and the polarization of the incident light ($e_i$), which was propagating in the Z direction. The zero-angle direction is defined by the experimental system as a reference at the beginning of the experiments. We found that the Raman scattering intensities of all three modes: $B_{3g}$, $A_g^1$, and $A_g^2$, are significantly dependent on the polarization angle of the excitation light (Fig. 2a). The Raman peak intensities of the $B_{3g}$, $A_g^1$, and $A_g^2$ modes are plotted as a function of angle $\theta_e$, as shown in Figs. 2b-2d, respectively, where $\theta_e$ stands for the sample rotation angle relative to the initial experimental 0° angle. The behaviors of polar plots are obviously different for the three different Raman modes, which represent direct fingerprints of the in-plane anisotropy of GeS. The polar plot of Raman intensities in both the $A_g^1$ and $A_g^2$ modes exhibits periodicity with an angle period of 180°.[6] In contrast, the polar plot of the $B_{3g}$ Raman mode shows a periodic variation with a period of 90°. In addition, there is a phase difference of 90° between the $A_g^1$ and $A_g^2$ polar plots which will be discussed in more detail. In the cross-polarization ($\perp$) configuration, the variation periods of Raman scattering intensity for all the three Raman modes are 90° (Figs. S2a-c). The intensity of the $A_g^1$ Raman mode corresponding to the lattice displacement along the armchair direction becomes the strongest, when the polarization of incident light is parallel to the same direction, and that of the $A_g^2$ Raman mode reaches to the lowest in the parallel-polarization configuration, as demonstrated by the Raman spectroscopy of other orthorhombic structures (e.g., BP and SnS).[5,6,20,21,26,29-31]

The Raman scattering intensity is proportional to $|e_i \cdot R \cdot e_s^T|^2$, where $R$ is the Raman tensor, and $e_i$ and $e_s$ are basic unit vectors of the incident and scattered light, respectively. The vector $e_i$ is described by the expression (0, cos$\theta$, sin$\theta$), where $\theta$ is the angle of the polarization of the incident laser relative to the armchair direction (Fig. S3), $e_s$ is described by (0, cos$\theta$, sin$\theta$) and (0, sin$\theta$, cos$\theta$) in the parallel and cross-polarized configurations, respectively. For the orthorhombic phase ($D_{2h}$ space group), the Raman tensors of the $A_g$ ($R(A_g)$) and $B_{3g}$ ($R(B_{3g})$) modes can be given as

$$R(B_{3g}) = \begin{pmatrix} 0 & 0 & 0 \\ 0 & 0 & k \\ 0 & k & 0 \end{pmatrix}$$

$$R(A_g) = \begin{pmatrix} l & 0 & 0 \\ 0 & m & 0 \\ 0 & 0 & n \end{pmatrix}, \quad (1)$$

where $l$, $m$, and $k$ are Raman tensor parameters. Then, the dependence of the Raman intensity, $I$, can be described as

$$I(A_g, //) = (m\cos^2\theta + n\sin^2\theta) \quad (2)$$
$$I(B_{3g}, //) = k^2(\sin 2\theta)^2 \quad (3)$$
$$I(A_g, \perp) = \frac{(n-m)^2}{4}(\sin 2\theta)^2 \quad (4)$$
$$I(B_{3g}, \perp) = k^2(\cos 2\theta)^2 \quad (5)$$

The experimental results shown in polar plots are well-reproduced by Eqs. (2)-(5) with an offset, as demonstrated by the solid curves in Figs. 2b-2d and S1a-1c. Typically, $m$ is larger (smaller) than $n$ for the $A_g^1$ ($A_g^2$) mode, which is consistent with the obtained fitting results. As a result, the experimental results for the $B_{3g}$ and $A_g$ modes can be explained by the lattice vibrations in the zigzag and armchair directions of GeS crystal, which suggests that polarized Raman scattering spectroscopy is an effective method for checking the crystal direction of layered GeS.[21, 29-31]

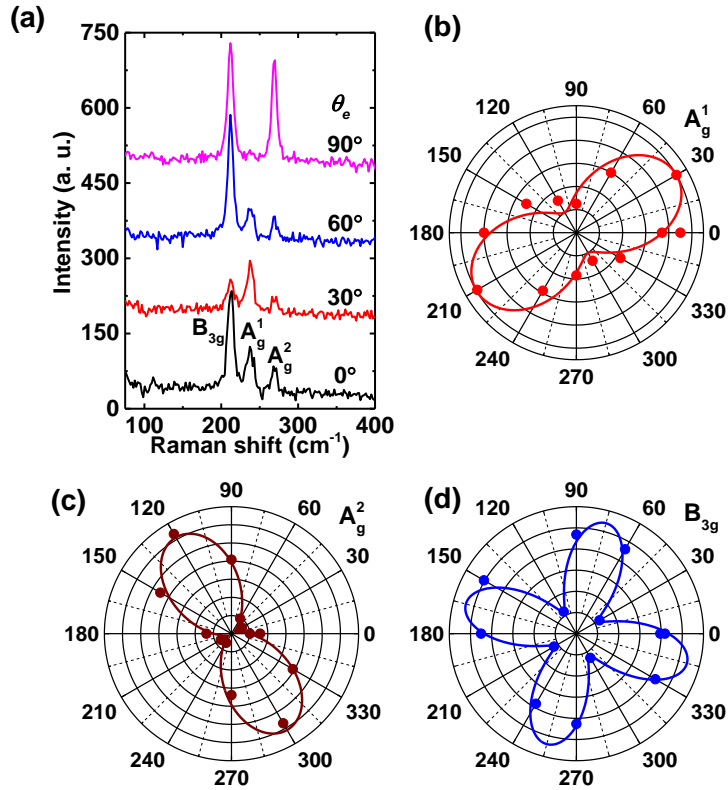

**Fig. 2.** (a) Polarized Raman scattering spectra of layered GeS at various angles in a parallel-polarization (//) configuration. The zero angle direction is defined by the experimental system as a reference at the beginning of the experiment. (b), (c), and (d) Polar plots of Raman peak intensities of the $A_g^1$, $A_g^2$, and $B_{3g}$ modes, respectively, as a functional of excitation polarization direction $\theta_e$. The red curves in the polar plots are fitting results using Eq. (2) for (b) and (c) and (3) for (d), plus an offset.

We have investigated the optical absorption and PL spectra to probe the electronic states of layered GeS. Fig. 3a (black curve) shows the PL spectrum in a 230 nm-thick GeS flake. The PL spectrum shows a clear peak around 1.66 eV under the weak excitation conditions of ~1.4 kW/cm$^2$. The inset of Fig. 3a shows the PL intensity image of the GeS flake monitored from 1.4 to 2.0 eV, in which clear light emission from the GeS can be observed. We also measured PL spectra with varying the excitation power density to obtain insight into the PL properties (Fig. S4). The PL intensity shows a simple linear dependence on the excitation power without saturating, which suggests that the PL originates from intrinsic states and not impurities or vacancy states.[6,32]

We also measured the optical absorption spectrum of GeS to reveal the electronic structure and mechanism of the PL, as shown in Fig. 3a (red curve). The absorption coefficient (α) begins to increase at around 1.6 eV and reaches a relatively large value of 1.2 × 10$^5$ cm$^{-1}$ at around 2.0 eV, which indicates that the optical absorption edge is around 1.6 eV. From the large value of the absorption coefficient, the optical absorption above 1.6 eV comes from an optically allowed transition. The small negative value of absorption coefficient below 1.6 eV in the transparent region can be observed, which results from the optical interferences in the transmission measurements.[33,34] The experimental value of the optical absorption edge is consistent with the previously reported optical absorption measurements of bulk GeS.[15,19]

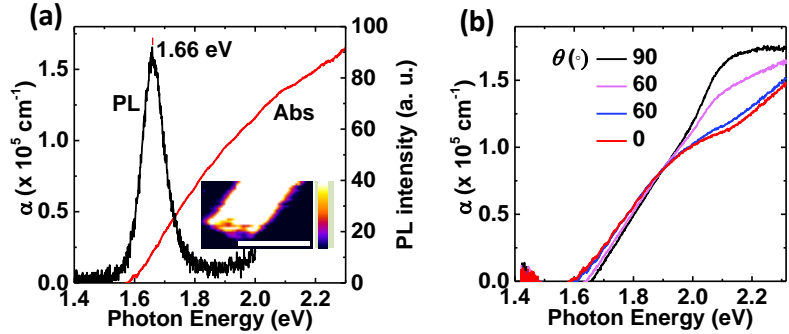

**Fig. 3.** (a) Optical absorption and PL spectra of a 230 nm-thick GeS flake. The inset shows PL image monitored from 1.4 to 2.0 eV. The scale bar in the image is 10 μm. (b) Polarization dependence of optical absorption spectra for various polarization angles, θ, relative to the armchair direction.

The Raman scattering spectra confirm the anisotropic structure of layered GeS, which may lead to its anisotropic electronic properties.[15] Next, polarized absorption spectroscopy was performed to address the anisotropic electronic properties experimentally. Before performing this experiment, we used polarized Raman scattering spectroscopy to clarify the crystal direction of the GeS flake (thickness 100 nm). Fig. 3b shows the polarized absorption spectra of the GeS flake.[1,35] The absorption spectrum clearly changes depending on the polarization direction of the incident light. The absorption onsets corresponding to the optical absorption edge in the armchair (θ = 0°) and zigzag (θ = 90°) orientations are 1.58 and 1.66 eV, respectively. The experimental results clearly show that the optical absorption properties show significant anisotropy depending on the crystal direction, and the optical absorption edge occurs at lower energy side in the armchair direction than in the zigzag direction.

First-principles DFT calculations were carried out to check the electronic band structure of layered GeS. Fig. 4a shows the calculated band dispersion of bulk GeS. In the calculated band structure, the conduction band bottom is at the Γ point, and the valence band top at the Γ point is very close to that near the X point, which results in a small energy difference of ~0.03 eV between the direct optical transition (Γ-Γ line, transition I indicated by the green arrow), and the indirect one (X-Γ line, transition II indicated by the green arrow).[13,15, 36] The DFT calculation predicts that bulk GeS is an indirect band gap semiconductor. However, the energy difference between the indirect and direct optical transitions is very small and comparable to the thermal energy at room temperature (26 meV).

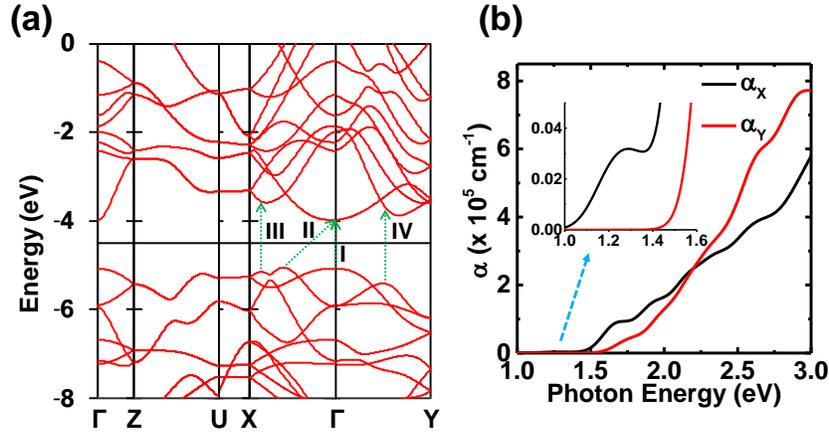

**Fig. 4.** (a) Electronic band structure of bulk GeS obtained by DFT calculations. (b) Calculated absorption coefficients for GeS along the armchair (X, $\theta = 0°$) and zigzag (Y, $\theta = 90°$) directions, denoted as $\alpha_X$ and $\alpha_Y$, respectively. Arrow I and II correspond to the optical transition between the direct optical transition (Γ-Γ line) and the indirect one (X-Γ line), respectively. Arrow III and IV correspond to the optical transition along the X and Y directions, respectively. The inset shows the absorption coefficient spectra expanded in the range of 1.0-1.6 eV.

The absorption coefficient ($\alpha$) spectra were also calculated by DFT using the obtained dielectric constants along the armchair (X) and zigzag (Y) directions, denoted as $\alpha_X$ and $\alpha_Y$, respectively. Fig. 4b shows the calculated $\alpha_X$ and $\alpha_Y$ absorption spectra of bulk GeS. According to the results of the previous DFT calculation, the effect of the Coulomb interaction in bulk GeS is evaluated as very small (< 10 meV), and the Coulomb interaction can be neglected in this calculation.[9] The very weak optical absorption around 1.26 eV is observed due to the partially dipole-allowed transition between Γ-Γ line of the momentum space in the calculated absorption spectrum (inset of Fig. 4b). The calculated $\alpha_X$ with the absorption edge at around 1.5 eV corresponding to the transition III (green arrow) in Fig. 4(a) is consistent with the experimentally obtained value in Fig. 3a. In contrast, the absorption in the Y direction ($\alpha_Y$) corresponding to the transition IV (green arrow) in Fig. 4a starts from around 1.6 eV, and the absorption intensity is weak compared with $\alpha_X$ in the same energy region. These calculated results are in agreement with the experimentally observed results that the optical absorption edge in the armchair direction is lower than that in the zigzag direction and that the absorption coefficient in the armchair direction is larger than that in the zigzag direction, as shown in Fig. 4b. These results verify that the electronic structure and corresponding optical absorption properties are anisotropic near the optical band edge of GeS. Moreover, the experimental results show that the energy of the PL peak corresponds well to the optical absorption edge along the X direction due to the optical transition shown in Fig. 3a, which also suggests that the PL mainly arises from the direct optical transition of GeS along the armchair direction.

The polarization dependence of the PL spectrum was measured to reveal the anisotropy of the electronic structure of layered GeS. Fig. 5a shows the PL spectra of a 170 nm-thickness GeS flake, obtained by varying the polarization angle of the excitation light $\theta$. No significant PL peak shift can be observed, similar to the case of BP.[9] The intensity of the PL spectra varies depending on the polarization angle of the excitation light. Fig. 5b shows a polar plot of PL intensity as a function of $\theta$. The experimental data in the polar plot can be fitted by a $\cos^2\theta$ function,[8] as indicated by the solid line in the plot. These results verify the anisotropic characteristics of the PL of layered GeS. The anisotropic PL signal is characterized by the linear polarization degree ($P$), defined as follows

$$p = (I_A - I_Z)/(I_A + I_Z) \quad (6)$$

where $I_A$ and $I_Z$ are the PL intensities with the polarization of the excitation light parallel to the armchair and zigzag direction, respectively.[3] The calculated value of $P$ is about 0.21 at an excitation photon energy of 2.33 eV. This similar behavior is also confirmed by the observation of similar results in another sample (Fig. S6) with a consistent $P$ of 0.20.

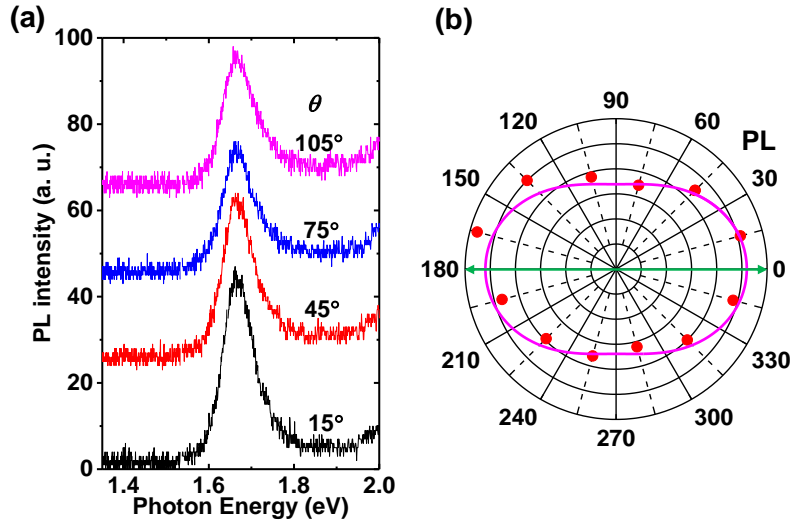

**Fig. 5.** (a) Excitation polarization dependence of PL spectra for $\theta$ of 15, 45, 75, and 105°. (b) Polar plots of PL intensity as a function of $\theta$. The armchair direction (green arrow line) was determined by the polarized Raman spectroscopy, as shown in Fig. S5 ($A_g^1$ mode).

The optical excitation and emission processes will contribute to the anisotropic PL properties of GeS.[6,37] The experimentally obtained absorption coefficient spectra in Figure 3b show that the optical absorption coefficient ($\alpha_X = 1.5 \times 10^5$ cm$^{-1}$) for the armchair direction ($\theta = 0°$) is slightly smaller than that in the zigzag direction ($\alpha_Y = 1.8 \times 10^5$ cm$^{-1}$, $\theta = 90°$) with a very small difference at 2.33 eV, which implies that the excitation process could not be responsible for relatively high $P$ in the polarization dependence of the PL spectra. In contrast, the optical transition at the optical band edge (around 1.66 eV) shows a significant relative difference between the armchair ($\theta = 0°$) and zigzag ($\theta = 90°$) directions, and larger absorption intensity is observed in the armchair direction than in the zigzag direction, as shown in Figure 3b. These experimental results are also consistent with the absorption spectra obtained by the DFT calculations (Fig. 4b). Thus, the experimentally observed anisotropic PL mainly comes from the emission process due to the anisotropic band structure along the zigzag and armchair directions of GeS.

## CONCLUSIONS

In summary, the anisotropic optical properties of layered GeS were studied using polarized Raman scattering, absorption and PL spectroscopies. Three Raman scattering peaks assigned to the $B_{3g}$, $A_g^1$, and $A_g^2$ modes with strong polarization dependence were revealed in the GeS flakes, and polarized Raman spectroscopy has been established to be a useful method to determine the crystal direction of GeS. PL with a peak at around 1.66 eV was observed at room temperature, which was attributed to the direct band gap transition on the basis of the experimental and DFT calculated optical absorption spectra. The PL also exhibits significant polarization dependence similar to that of the $A_g^1$ mode Raman scattering peak, revealing the anisotropic optical transition near the optical band edge, as confirmed by the anisotropic absorption behavior. The reintroduced anisotropic layered material GeS exhibits unique anisotropic structure and optical properties, suggesting that it has the highly potential for the fabrication of novel electronic and optical devices.

## EXPERIMENTAL SECTION:

GeS flakes were mechanically exfoliated from bulk single crystals and transferred onto typical 300 nm-thick $SiO_2$/Si or quartz substrates. The thickness was obtained from the height profile obtained by atomic-force microscopy (AFM). PL and polarized Raman spectroscopy were measured using Micro-Raman setups (Raman touch, Nanophoton and LabRAM HR Evolution, Horiba) with a semiconductor laser diode (2.33 eV). All measurements were conducted at room temperature in an air atmosphere. The typical excitation power was maintained under 20 μW. The optical absorption spectrum was measured from the transmission values and thickness of the sample using a halogen light source in the Raman microscope.[1,35] The polarization dependence of the absorption spectrum was measured using a linear polarizer in the optical path of the transmitted light.

**Corresponding Authors**
*E-mail: tan.dezhi.63m@st.kyoto-u.ac.jp

*E-mail: matsuda@iae.kyoto-u.ac.jp

ACKNOWLEDGMENT
This study was supported by a Grant-in-Aid for Grants-in-Aid for JSPS KAKENHI (Nos. 25400324, 24681031, 15K13500, 26107522, 25246010 and 15F15313) and by Precursory Research for Embryonic Science and Technology (PRESTO) from the Japan Science and Technology Agency (JST). The authors thank E. Iso, and Y. Nakata for support of polarized Raman spectroscopy.

(1) Yuan, H.; Liu, X.; Afshinmanesh, F.; Li, W.; Xu, G.; Sun, J.; Lian, B.; Curto, A. G.; Ye, G.; Hikita, Y.; Shen, Z.; Zhang, S. C.; Chen, X.; Brongersma, M.; Hwang, H. Y.; Cui, Y. Polarization-Sensitive Broadband Photodetector Using a Black Phosphorus Vertical P-N Junction. *Nat. Nanotechnol*. **2015**, 10, 707–713.
(2) Nemilentsau, A.; Low, T.; Hanson, G. Anisotropic 2D Materials for Tunable Hyperbolic Plasmonics. *Phys. Rev. Lett*. **2016**, 116, 066804.
(3) Son, Y.-W.; Cohen, M. L.; Louie, S. G. Half-metallic Graphene Nanoribbons. *Nature* **2006**, 444, 347–349.
(4) Mak, K. F.; Shan, J. Photonics and Optoelectronics of 2D Semiconductor Transition Metal Dichalcogenides. *Nat. Photon*. **2016**, 10, 216–226.
(5) Xia, F., Wang, H.; Jia, Y. Rediscovering Black Phosphorus as an Anisotropic Layered Material for Optoelectronics and Electronics. *Nat. Commun*. **2014**, 5, 4458.
(6) Wang, X.; Jones, A. M.; Seyler, K. L.; Tran, V.; Jia, Y.; Zhao, H.; Wang, H.; Yang, L.; Xu, X.; Xia, F. Highly Anisotropic and Robust Excitons in Monolayer Black Phosphorus. *Nat. Nanotech.* **2015**, 10, 517–521.
(7) Li, L.; Yu, Y.; Ye, G. J.; Ge, Q.; Ou, X.; Wu, H.; Feng, D.; Chen, X. H.; Zhang, Y. Black Phosphorus Field-Effect Transistors. *Nat. Nanotechnol*. **2014**, 9, 372–377.
(8) Lorchat, E.; Froehlicher, G.; Berciaud, S. Splitting of Interlayer Shear Modes and Photon Energy Dependent Anisotropic Raman Response in N‑Layer ReSe$_2$ and ReS$_2$. *ACS Nano* 2016, 10, 2752–2760.
(9) Shi, G.; Kioupakis, Emmanouil. Anisotropic Spin Transport and Strong Visible-Light Absorbance in Few-Layer SnSe and GeSe. *Nano Lett*. **2015**, 15, 6926−6931.
(10) Mannix, A. J.; Zhou, X.-F.; Kiraly, B.; Wood, J. D.; Alducin, D.; Myers, B. D.; Liu, X.; Fisher, B. L.; Santiago, U.; Guest, J. R.; Yacaman, M. J.; Ponce, A.; Oganov, A. R.; Hersam, M. C.; Guisinger, N. P. Synthesis of Borophenes: Anisotropic, Two-dimensional Boron Polymorphs. *Science* **2015**, 350, 1513–1516.
(11) Lan, C.; Li, C.; Yin, Y.; Guo, H.; Wang, S. Synthesis of Single-Crystalline GeS Nanoribbons for High Sensitivity Visible-Light Photodetectors. *J. Mater. Chem. C* **2015**, 3, 8074–8079.
(12) Ulaganathan, R. K.; Lu, Y. -Y.; Kuo, C. -J.; Tamalampudi, S. R.; Sankar, R.; Boopathi, K. M.; Anand, A.; Yadav, K.; Mathew, R. J.; Liu, C. -R.; Choue, F. C.; Chen, Y. -T. High Photosensitivity and Broad Spectral Response of Multi-Layered Germanium Sulfide Transistors. *Nanoscale* **2016**, 8, 2284–2292.
(13) Gomes, L. C.; Carvalho, A. Phosphorene Analogues: Isoelectronic Two-dimensional Group-IV Monochalcogenides with Orthorhombic Structure. *Phys. Rev. B* **2015**, 92, 085406.
(14) Gomes, L. C.; Carvalho, A.; Castro Neto, A. H. Enhanced Piezoelectricity and Modified Dielectric Screening of Two-dimensional Group-IV Monochalcogenides. *Phys. Rev. B* **2015**, 92, 214103.
(15) Makinistian, L.; Albanesi, E. A. First-Principles Calculations of the Band Gap and Optical Properties of Germanium Sulfide. *Phys. Rev. B* **2006**, 74, 045206.
(16) Tuttle, B. R.; Alhassan, S. M.; Pantelides, S. T. Large Excitonic Effects in Group-IV Sulfide Monolayers. *Phys. Rev. B* **2015**, 92, 235405.
( 1 7) Li, F.; Liu, X.; Wang, Y.; Li, Y. Germanium Monosulfide Monolayer: A Novel Two-dimensional Semiconductor with A High Carrier Mobility. *J. Mater. Chem. C* **2016**, 4, 2155–2159.
(18) Gomes, L. C.; Carvalho, A.; Castro Neto, A. H. Vacancies and Oxidation of 2D Group-IV Monochalcogenides. arXiv:1604.04092.
(19) Vaughn, D. D.; Patel, R. J.; Hickner, M. A.; Schaak, R. E. Single-Crystal Colloidal Nanosheets of GeS and GeSe. *J. Am. Chem. Soc*. **2010**, 132, 15170–15172.

(20) Wu, J.; Mao, N.; Xie, L.; Xu, H.; Zhang, J. Identifying the Crystalline Orientation of Black Phosphorus Using Angle-Resolved Polarized Raman Spectroscopy. *Angew. Chem. Int. Ed.* **2015**, 54, 2366–2369.
(21) Ribeiro, H. B.; Pimenta, M. A.; de Matos, C. J. S.; Moreira, R. L.; Rodin, A. S.; Zapata, J. D.; de Souza, E. A. T.; Neto, A. H. C. Unusual Angular Dependence of the Raman Response in Black Phosphorus. *ACS Nano* **2015**, 9, 4270–4276.
(22) Ferrari, A. C.; Basko, D. M. Raman Spectroscopy as A Versatile Tool for Studying the Properties of Graphene. *Nat. Nanotechnol*. **2013**, 8, 235–246.
(23) Zhang, X.; Tan, Q. -H.; Wu, J. -B.; Shi, W.; Tan, P. -H. Review on the Raman Spectroscopy of Different Types of Layered Materials. *Nanoscale* **2016**, 8, 6435–6450.
(24) Luo, X.; Lu, X.; Koon, G. K. W.; Castro Neto, A. H.; Özyilmaz, B.; Xiong, Q.; Quek, S. Y. Large Frequency Change with Thickness in Interlayer Breathing Mode-Significant Interlayer Interactions in Few Layer Black Phosphorus. *Nano Lett*. **2015**, 15, 3931–3938.
(25) Wolverson, D.; Crampin, S.; Kazemi, A. S.; Ilie, A.; Bending, S. J. Raman Spectra of Monolayer, Few-Layer, and Bulk ReSe2: An Anisotropic Layered Semiconductor. *ACS Nano* **2014**, 8, 11154–11164.
(26) Wiley, J. D.; Buckel, W. J.; Schmidt, R. L. Infrared Reflectivity and Raman Scattering in GeS. *Phys. Rev. B* **1976**, 63, 2489–2496.
(27) Hsueh, H. C.; Warren, M. C.; Vass, H.; Ackland, G. J.; Clark, S. J.; Crain, J. Vibrational Properties of the Layered Semiconductor Germanium Sulfide under Hydrostatic Pressure: Theory and Experiment. *Phys. Rev. B* **1996**, 53, 14806–14817.
(28) Chandrasekhar, H. R.; Humphreys, R. G.; Card, M. Pressure Dependence of the Raman Spectra of the IV-VI Layer Compounds GeS and GeSe. *Phys. Rev. B* **1977**, 16, 2981.
(29) Kim, J.; Lee, J. -U.; Lee, J.; Park, H. J.; Lee, Z.; Lee, C.; Cheong, H. Anomalous Polarization Dependence of Raman Scattering and Crystallographic Orientation of Black Phosphorus. *Nanoscale* **2015**, 7, 18708–18715.
(30) Ling, X.; Liang, L.; Huang, S.; Puretzky, A. A.; Geohegan, D. B.; Sumpter, B. G.; Kong, J.; Meunier, V.; Dresselhaus, M. S. Low-Frequency Interlayer Breathing Modes in Few-Layer Black Phosphorus. *Nano Lett.* **2015**, 15, 4080–4088.
(31) Xia, J.; Li, X. -Ze.; Huang, X.; Mao, N.; Zhu, D. -D.; Wang, L.; Xu, H.; Meng, X. -M. Physical Vapor Deposition Synthesis of Two-dimensional Orthorhombic SnS Flakes with Strong Angle/Temperature-Dependent Raman Responses. *Nanoscale* **2016**, 8, 2063–2070.
(32) Jones, A. M.; Yu, H.; Ghimire, N. J.; Wu, S.; Aivazian, G.; Ross, J. S.; Zhao, B.; Yan, J.; Mandrus, D. G.; Xiao, D.; Yao, W.; Xu, X. Optical Generation of Excitonic Valley Coherence in Monolayer WSe2. *Nat. Nanotechnol*. **2013**, 8, 634–638.
(33) Eakin, R. T. Interference in Transmission Spectra of Overlapping Lorentzian Lines. *J. Quant. Spectrosc. Radiat. Transfer* **1988**, 39, 225–236.
(34) Süleymanov, R. A.; Ellialtíoğlu, Ş.; Akínoğlu, B. G. Layered semiconductor GeS as A Birefringent Stratified Medium. *Phys. Rev. B* **1995**, 52, 7806.
(35) Ling, X.; Huang, S.; Hasdeo, E. H.; Liang, L.; Parkin, W. M.; Tatsumi, Y.; Nugraha, A. R. T.; Puretzky, A. A.; Das, P. M.; Sumpter, B. G.; Geohegan, D. B.; Kong, J.; Saito, R.; Drndic, M.; Meunier, V.; Dresselhaus, M. S. Anisotropic Electron-Photon and Electron-Phonon Interactions in Black Phosphorus. *Nano Lett.* **2016**, 16, 2260–2267.
(36) Malone, B. D.; Kaxiras, E. Quasiparticle Band Structures and Interface Physics of SnS and GeS. *Phys. Rev. B* **2013**, 87, 245312.
(37) Zhang, S.; Yang, J.; Xu, R.; Wang, F.; Li, W.; Ghufran, M.; Zhang, Y. -W.; Yu, Z.; Zhang, G.; Qin, Q.; Lu, Y. Extraordinary Photoluminescence and Strong

Temperature/Angle-Dependent Raman Responses in Few-Layer Phosphorene. *ACS Nano* **2014**, 8, 9590–9596.